# Rigorous analysis of highly tunable cylindrical Transverse Magnetic mode re-entrant cavities

J-M. Le Floch`,[1,2a], Y. Fan[1,2], M. Aubourg[3], D. Cros[3], N.C. Carvalho[1,2], Q. Shan[1,4], J. Bourhill[1,2], E.N. Ivanov[1,2], G. Humbert[3], V. Madrangeas[3], M.E. Tobar[1,2]

1. School of Physics, University of Western Australia, 35 Stirling Hwy, 6009 Crawley, Western Australia
2. ARC Centre of Excellence, Engineered Quantum Systems (EQuS, 35 Stirling Hwy, 6009 Crawley, Western Australia)
3. XLIM, UMR CNRS No 6172, 123 av. A. Thomas, 87060 Limoges Cedex, France
4. College of Mechatronics and Automation, National University of Defense Technology, Changsha, 410073, China

Cylindrical re-entrant cavities are unique three-dimensional structures that resonate with their electric and magnetic fields in separate parts of the cavity. To further understand these devices, we undertake rigorous analysis of the properties of the resonance using "in-house" developed Finite Element Method (FEM) software capable of dealing with small gap structures of extreme aspect ratio. Comparisons between the FEM method and experiments are consistent and we illustrate where predictions using established lumped element models work well and where they are limited. With the aid of the modeling we design a highly tunable cavity that can be tuned from 2 GHz to 22 GHz just by inserting a post into a fixed dimensioned cylindrical cavity. We show this is possible as the mode structure transforms from a re-entrant mode during the tuning process to a standard cylindrical Transverse Magnetic (TM) mode.

## I. INTRODUCTION

Microwave re-entrant cavities have been studied over 50 years for a variety of applications. The development of accurate structure modeling based on lumped element analysis [1-7] has made them very valuable for creating small volume high-Q resonators for filtering applications from room to cryogenic temperatures in comparison with large Whispering Gallery and Bragg mode resonant cavities [6-11]. Re-entrant cavities have also been used for producing displacement sensors for gravitational bar detectors [12-17], oscillators using Gunn diodes [18] and electron beam tubes [18], and for millimetre wave resonators [20-21]. Furthermore, the re-entrant cavity has a high frequency tuning capability, which make them very attractive for telecommunication systems [22-25] and characterizing dielectric materials as a function of frequency [26-27]. The electromagnetic field pattern of the re-entrant mode presents a very high capacitance between the internal post and the lid of the cavity. This high electric field confinement within the gap has been of some interest in plasma physics for developing high power electron beam microwave tubes. Mostly using rectangular re-entrant cavities it has been possible to generate output power of a few kWatts. For electron beam device applications, the design of klystron cavities (often at lower frequencies) use standard lumped circuit models adjusted to include plasma-wave effects [28-32]. Cylindrical re-entrant cavities have also been used for studying absorption processes in small liquid and solid materials [33-34], and also the breakdown of gases to E-field [35].

The microwave re-entrant cavity's high frequency selectivity to gap size fluctuations makes it also a unique device, capable of being developed for mechanical transducer applications [15-17, 36] and for investigating the dynamical Casimir Effect [37]. The re-entrant cavity could also prove to be a useful tool for cavity based searches for axions [38] and axion-like particles, where the ability to tune over a large frequency range while maintaining low electrical losses and high E-field intensity would enable sensitive experiments that cover a wide region of parameter space.

Corresponding author: Jean-Michel Le Floch; jeanmichel.lefloch@uwa.edu.au

Nevertheless, throughout all this research there is a lack of understanding of rigorous analysis on the mode structure as the position of the post inside the cavity widely varies to the regime where the approximations no longer hold. For example, lumped element models are only valid under certain approximations [39-43]. In this paper we undertake rigorous analysis using the Finite Element Method (FEM) and compare with the lumped element model and experimental results. First we compare results in the region where the approximation is valid, and then we construct a highly tunable cavity, which goes beyond the simple lumped model and can be tuned from 2 to 22 GHz. The lower frequency of 2GHz corresponds to a 3μm gap size achieved with a macroscopic tuning apparatus. It is shown that the mode structure transforms from a re-entrant mode during the tuning process to a standard cylindrical Transverse Magnetic (TM) mode by the time the post is finally removed. We also explain with the aid of field plots why the LCR model fails with very large gap sizes.

## II. COMPUTATIONAL AND EXPERIMENTAL RESULTS

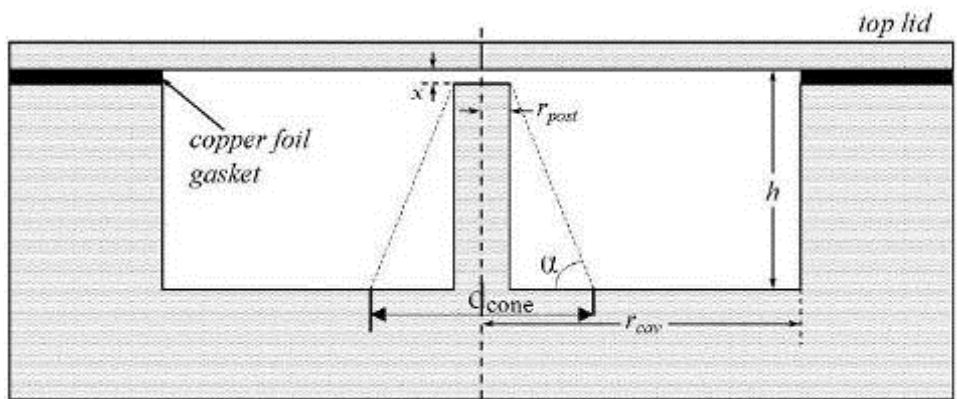

Fig. 1. Cross-section of the *r*, *z* plane of a cylindrical re-entrant cavity of height *h* and radius $r_{cav}$, with a metal post in the centre. Here α denotes the angle of the post, $d_{cone}$ the diameter of the base of the post, $r_{post}$ the radius of the tip and *x* defines the gap between the tip and the lid.

General design and modeling of re-entrant cavities requires the calculation of the resonance frequency, *f*, with respect to the gap spacing *x* and the sensitivity *df/dx*. In Fig 1, we present the general topology with a conical post [2, 3]. For high sensitivity metrological applications the figure of merit to optimize is $\frac{GF}{f}\left(\frac{df}{dx}\right)$, which allows maximum transductance of displacement to electrical energy [25]. The factor of merit has been chosen as such it is independent from the resonance frequency, metal cavity losses, and just dependent from the frequency sensitivity with the gap size (df/dx). It is then applicable to any cavity resonance sizes and any metal enclosure for a given gap size.

The geometric factor *(GF)* is a parameter, which relates the resonant modes coupling to the metallic losses of the material, and is given by the following formula (here $Q_0$ is the unloaded quality factor and *Rs* is the surface resistance):

$$GF = \omega_{res} \frac{\iiint\limits_{\text{Total Volume}} \mu_o |H|^2 dV}{\iint\limits_{\text{Surface}} |H_{tangential}|^2 dS} = Q_0 Rs \quad (1)$$

Note that GF only depends on the mode field pattern, and is in the units Ohms.

Using finite element electromagnetic analysis for various topologies, we have calculated the sensitivity for gap sizes of order a nanometer to a millimeter, which are shown in fig. 3. The calculations have been conducted using "in-house" software developed at XLIM institute in France, dedicated to microwave resonator design. The fringing fields near the gap are difficult to model accurately using numerical methods. The extreme aspect ratio of the gap spacing to cavity height is very

Corresponding author: Jean-Michel Le Floch; jeanmichel.lefloch@uwa.edu.au

challenging for mesh creation as the ratio can be greater than 100. It is usually necessary to use very large numbers of mesh cells to obtain good accuracy and the computational time can then become very large. This XLIM software enables us to perform the simulations where typical commercial simulation software fail when the ratio of the gap to the cavity height becomes too large. It is then necessary to mesh differently depending of this dimensional ratio with respect to the resonance frequency. We used non-equidistant weighted points triangular mesh cell where the field is maximum to enhance the precision of the calculations. It is then optimized with a freeware gmsh. The finer the mesh from nano to micrometer gaps, the bigger the required calculation power required. This is made available with the CALI (Calcul en Limousin) server. To give consistent results at extra small gaps, we implemented a thousand of wavelength mesh cell to discretize the gap. This requires a few tens of Gb of memory to conduct the calculations even though all symmetries of the structure are used. This obviously needs a lot of time for computing the solutions with very large matrices. An example of such a mesh is shown in fig. 2.

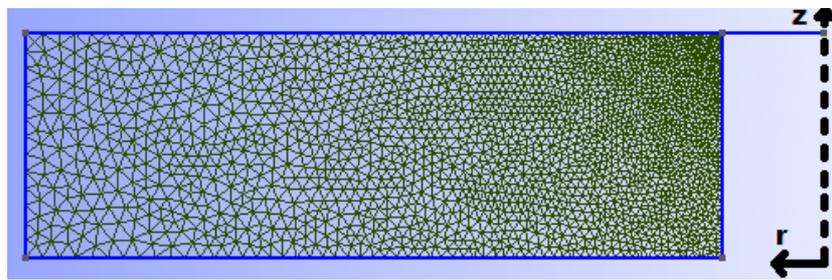

Fig. 2. Plot of mesh for half of the reentrant cavity gap structure using revolution symmetry, which is used for calculating frequencies on the order of nanometer gap sizes

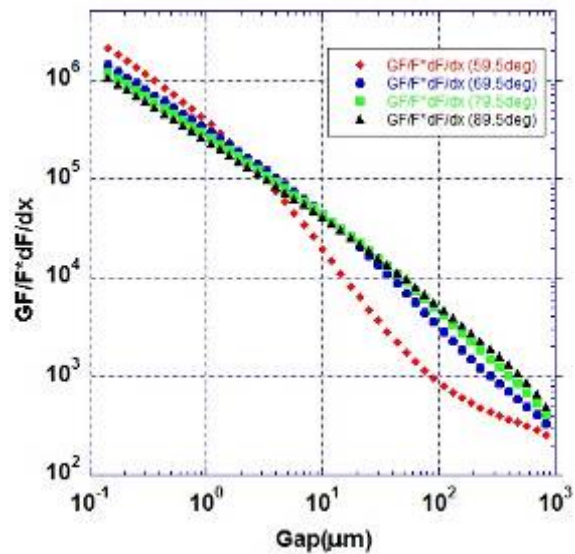

Fig 3: Calculation of $\frac{GF}{f}\left(\frac{df}{dx}\right)$ using finite element analysis for various angles α (59.5 to 89.5 degrees) as shown in figure 1. Below 30 μm the sensitivities are very similar and only vary by up to a factor of two and sub μm gap sizes. Thus, for simplicity we report experimentally on cavities with α = 90 degrees.

*A. Lumped model*

Figure 4 shows the equivalent model of the re-entrant cavity following the conventional process for any resonant structure in the microwave regime. A resonant cavity can be represented with an equivalent LCR circuit where LC is related to frequency and R gives information on the losses. In the case of a re-entrant cavity, we have an additional capacitance, which models the field discontinuity, i.e. the gap formed by the post and the top lid.

Corresponding author: Jean-Michel Le Floch; jeanmichel.lefloch@uwa.edu.au

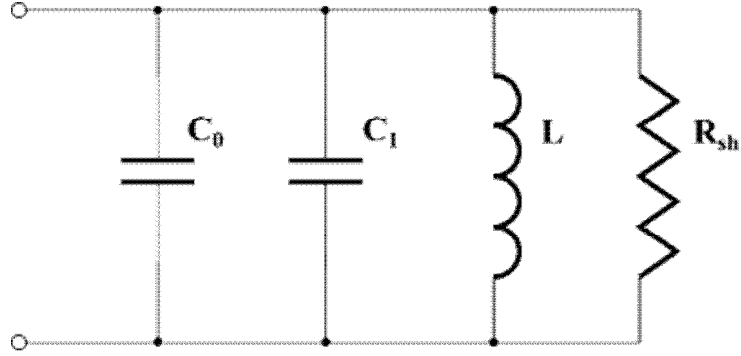

Fig 4: Equivalent lumped element circuit of the fundamental re-entrant cavity mode.

The formula for approximating the resonance frequency of the fundamental re-entrant cavity mode [3] is given below in equation (2):

$$L = \mu_0 \frac{h}{2\pi} \ln\left(\frac{r_{cav}}{r_{post}}\right) \quad (2a)$$

$$C_0 = \varepsilon_0 \frac{\pi r_{post}^2}{d} \quad (2b)$$

$$C_1 = 4\varepsilon_0 r_{post} \ln\left(\frac{e\sqrt{(r_{cav}-r_{post})^2 + h^2}}{2d}\right) \quad (2c)$$

$$R_{sh} = \frac{2\pi\omega^2 L^2}{R_{surface}\left(\frac{h-d}{r_{post}} + \frac{h}{r_{cav}} + 2\ln\left(\frac{r_{cav}}{r_{post}}\right)\right)} \quad (2d)$$

$$R_{Surface} = \frac{1}{\sigma\delta} \text{ and } \delta = \sqrt{\frac{2}{\omega\mu_0\sigma}} \quad (2e)$$

$$f = \frac{\omega}{2\pi} = \frac{1}{2\pi\sqrt{r_{post} h\left(\frac{r_{post}}{2d} + \frac{2}{\pi}\ln\left(\frac{e\sqrt{(r_{cav}-r_{post})^2 + h^2}}{2d}\right)\right) \ln\left(\frac{r_{cav}}{r_{post}}\right)}} \quad (2f)$$

Here $\mu_0$, $\varepsilon_0$ are the permeability and the permittivity of free space respectively, $\sigma$ is the conductivity of the metal conductor and $\delta$ represents the skin depth.

The corresponding *GF* using the lumped circuit analysis [19] is given by equation (3):

$$GF = \omega_0 \mu_0 \frac{h \ln\left(\frac{r_{cav}}{r_{post}}\right)}{\frac{h}{r_{post}} + \frac{h}{r_{cav}} + 2\ln\left(\frac{r_{cav}}{r_{post}}\right)} \quad (3)$$

Thus, the Q-factor may be calculated knowing the surface resistance of the metal by combing with equation (1).

### B. Comparison between modeling and experiments

Measurements have been undertaken and compared with the finite element analysis and the lumped element model [3-4]. Various copper foil gaskets (as shown in figure 5) of a few micrometers thick (with 15% uncertainty) were used to vary the gap spacing by placing the gaskets between the lid of the cavity and the cylindrical wall. The measurements were conducted using a vector network analyzer with

Corresponding author: Jean-Michel Le Floch; jeanmichel.lefloch@uwa.edu.au

a magnetic probe inserted through the side of the cavity in a one-port configuration to couple to the azimuthal magnetic field component. The resonant frequency and unloaded Q-factor were derived from the measured reflection coefficient $S_{11}$ [44-45]. Due consideration was given to frequency pulling and external losses induced by the loop probe, by measuring the resonant frequency and the Q-factor of the cavity as a function of the probe position. In this way we made sure measurements were taken in the under coupled regime where the influence of the probe is minimized and we measure the intrinsic properties of the resonator. This thorough procedure ensures the coupling of the field with the probe is negligible.

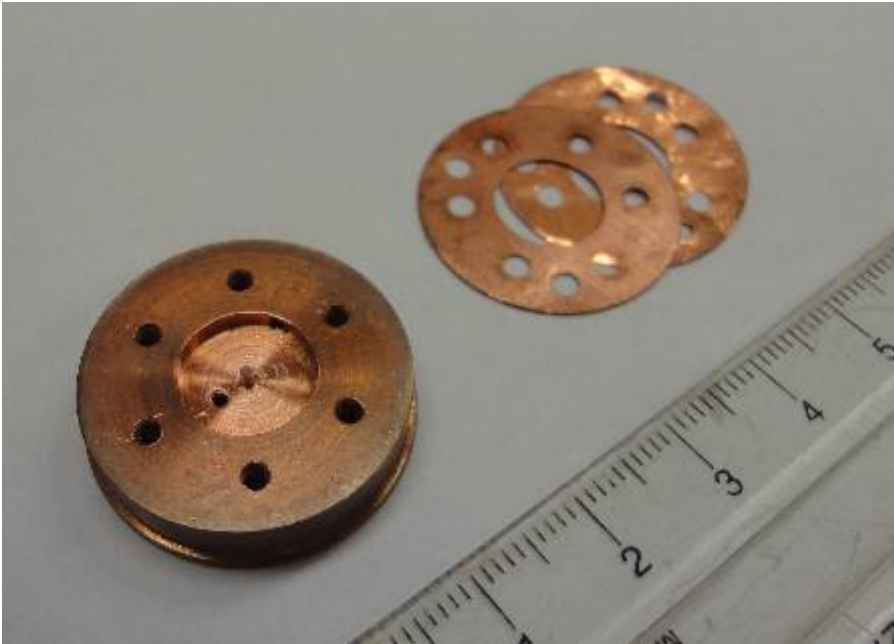

Fig. 5. Photograph of the copper cavity and gaskets for the re-entrant cavity under investigation.

The corresponding plots in figures 6 compare the measured and calculated results of a cylindrical cavity with radius $r_{cav}$ = 4.97 mm, height $h$ = 1.4 mm, and $r_{post}$ = 0.5 mm.

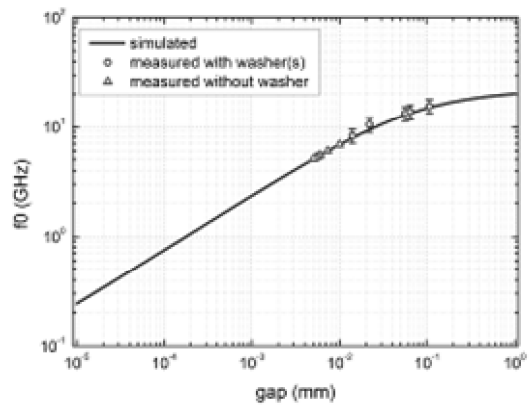

(a)

Corresponding author: Jean-Michel Le Floch; jeanmichel.lefloch@uwa.edu.au

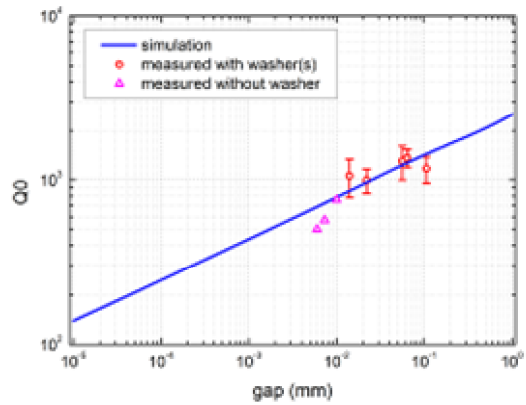

(b)

Fig. 6. Fundamental re-entrant mode frequency and Q-factor as a function of gap spacing (*x*), assuming an electric conductivity of $3.3.10^7$ S/m for copper. Measurements are plotted as points and are compared with the simulated Finite Element results (solid line).

The error-bars in the measured frequency data are obtained from the thickness tolerance (±15%) of the copper foil. The triangle data points in figure 6 are measured by calibrating the gap spacing from the frequency measurements of the circled data points.

TABLE I COMPARISON BETWEEN MEASURED DATA AND (LCR AND FEM) SIMULATION RESULTS WITH AN ELECTRIC CONDUCTIVITY OF $3.3.10^7$ S/M GIVEN WITH THE CONDUCTED MEASUREMENTS FIG. 6B.

| Gap simulated μm | $f_0$ GHz | $Q_0$ 10% error | $f_{sim}$ GHz | Q simulated | $f_{LCR}$ GHz | $Q_{LCR}$ |
|---|---|---|---|---|---|---|
| 10 | 6.886 | 760 | 6.878 | 787.3 | 6.949 | 802.7 |
| 7.311 | 6.006 | 570 | 6.003 | 725.9 | 6.049 | 747.5 |
| 5.97 | 5.476 | 502 | 5.486 | 689.1 | 5.518 | 713.3 |
| 5.546 | 5.298 | -* | 5.307 | 676.2 | 5.334 | 701.2 |
| 5.445 | 5.249 | -* | 5.263 | 672.9 | 5.289 | 698.2 |
| 5.058 | 5.11 | -* | 5.09 | 660.4 | 5.113 | 686.2 |

* Modes with very weak signals, so no Q-factors are calculated.

Measurement and computational frequency results for the re-entrant cavity at 296.5 K are given in Table I. Resonant frequency and Q-factor computations were performed for gap sizes between 1nm and 1mm. Measurements were conducted between 10 and 100μm. The differences between frequencies computed from the finite element analysis and those measured in experiment do not exceed more than 0.1%. Resonant frequencies measured from the network analyser are as precise as 0.001% and Q-factor values are approximately given to 10% error. Numerical accuracy is not a limiting factor in such a configuration.

The frequency and Q-Factor plots show very good consistency between the modeling with lumped element, Finite Element Analysis and the experiments. In this way we have confirmed the accuracy of our modeling. The difference in Q-factor values between simulations and measurement for very small gap is assumed to be directly coming from the copper oxidation. The oxidation is a very lossy dielectric material which could result in lowering the Q value.

Corresponding author: Jean-Michel Le Floch; jeanmichel.lefloch@uwa.edu.au

### III. DESIGN OF A TUNEABLE RE-ENTRANT CAVITY

A highly tunable microwave cavity was designed by implementing a fine threaded screw mechanism, which enabled a rotation-free translation of the metal pin as shown in figure 7. During translation a tight fit kept reasonable electrical contact of the pin to the cavity. In such a case we were able to reproduce previous measurements and reduce the gap size to as low as 3μm and as high as 1.4mm (empty cavity configuration) as shown on figure 8. The Hφ field component was excited by implementing a loop probe. The loop was positioned to stay in the under-coupled regime (no field disturbance from the probe) for measuring the unloaded Q-factor at each gap size. The 3μm-gap size limitation comes from the pin and the top lid not being parallel with the metal surface due to roughness from lathing and milling.

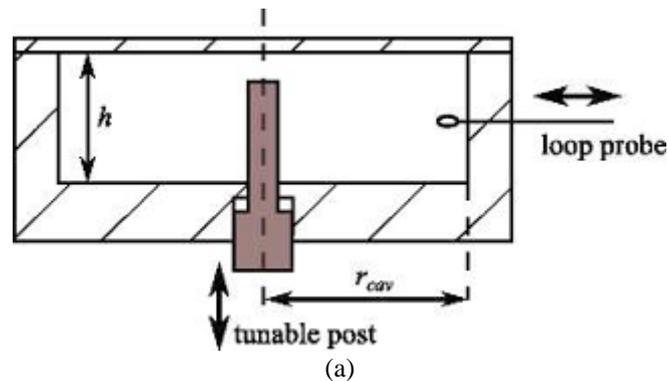

(a)

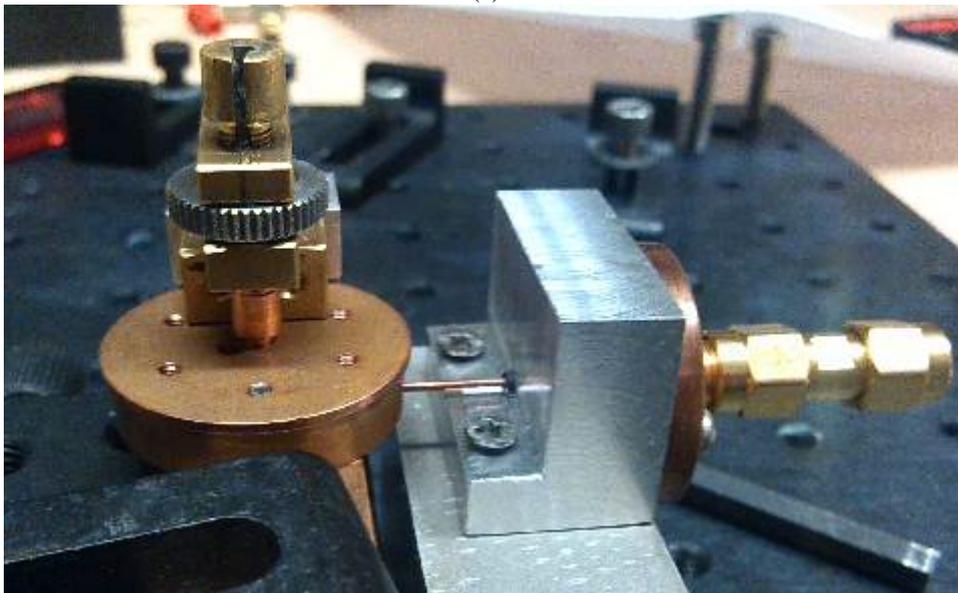

(b)

Figure 7: (a) Tunable cavity design with the rotation-free move pin, (b) Manufacturing of the ensemble mechanism.

A comparison between the equivalent lumped circuit model (LCR), the Finite Elements Method (FEM) and the measurements has been conducted and is illustrated in figure 8. It shows a good agreement between FEM and the measurements but highlights the limitation and inconsistency of the LCR model for a gap greater than 20 μm. This could be explained by the fact that the LCR equivalent circuit doesn't take into account the mode change from a re-entrant to a pure Transverse Magnetic (TM) mode.

Corresponding author: Jean-Michel Le Floch; jeanmichel.lefloch@uwa.edu.au

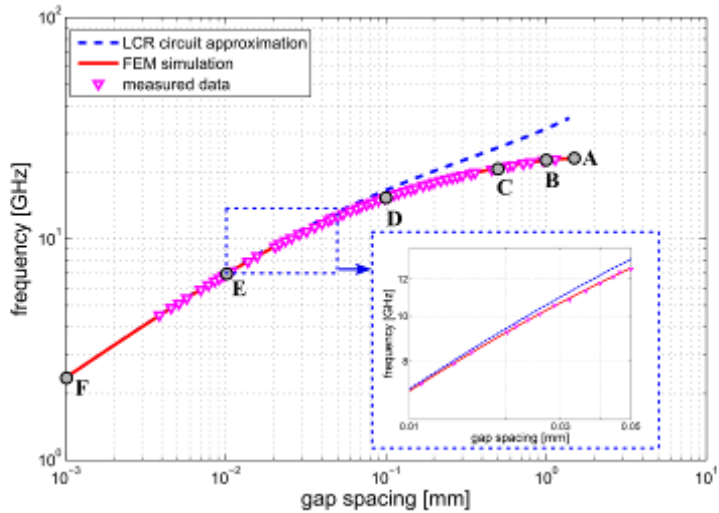

Figure 8: Frequency tuning with gap spacing comparison between the lumped elements analysis (LCR), the Finite Elements Method (FEM) and the measured data using the tuneable cavity. Points A to F show where a density plot of the mode has been calculated and is illustrated in figure 9.

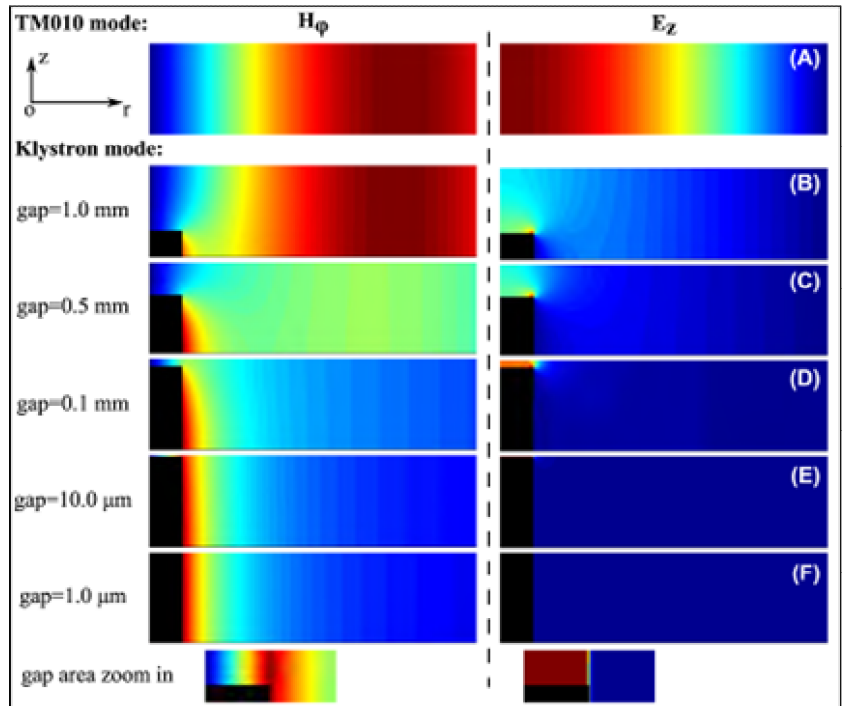

Figure 9: Density plot of the field pattern distribution in half of the cavity with gap size changing. The left column shows the azimuthal magnetic field component and the right column the axial electric field component. $H_\varphi$ and $E_z$ are the only components of the electromagnetic field distribution of the $TM_{010}$ and re-entrant modes. (A) to (F) refer to the gap to frequency change given previously on figure 8.

Figure 9 illustrates clearly how the field pattern changes from a gap size of 1 μm to all the way out. At 100 μm ((D) in figure 9) electric field fringing becomes significant and the magnetic field starts to be drawn away from the post. The lumped elements circuit model and the rigorous analysis starts to diverge at this point because the capacitance and inductance can no longer be only approximated as due to the gap and the post. However, for smaller gaps below 100 μm the model is as accurate as the finite element method, and much more easily to implement.

Corresponding author: Jean-Michel Le Floch; jeanmichel.lefloch@uwa.edu.au

We also characterize the tuneable re-entrant cavity in term of loss between the simulation and the measurements. All Q-factors are given for a very low coupling coefficient for which the values are the highest due to the minimization of probe losses. A comparison is made before and after the surface oxidation of copper is removed (surface cleaning). These results are plotted in figure 10.

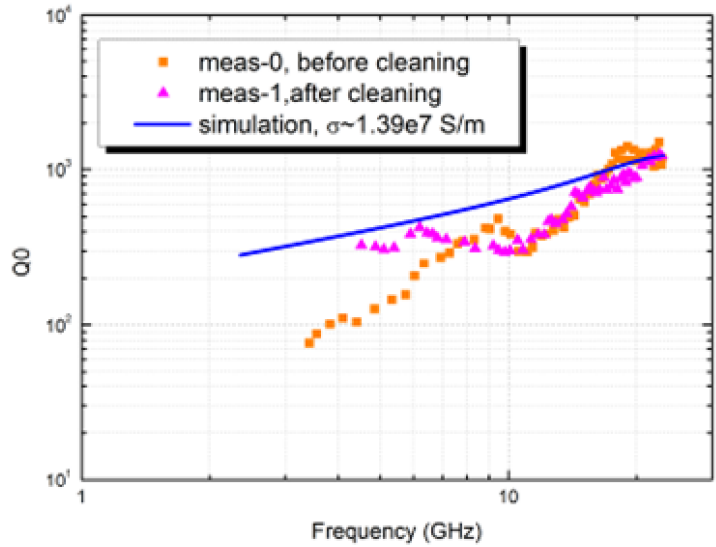

(a)

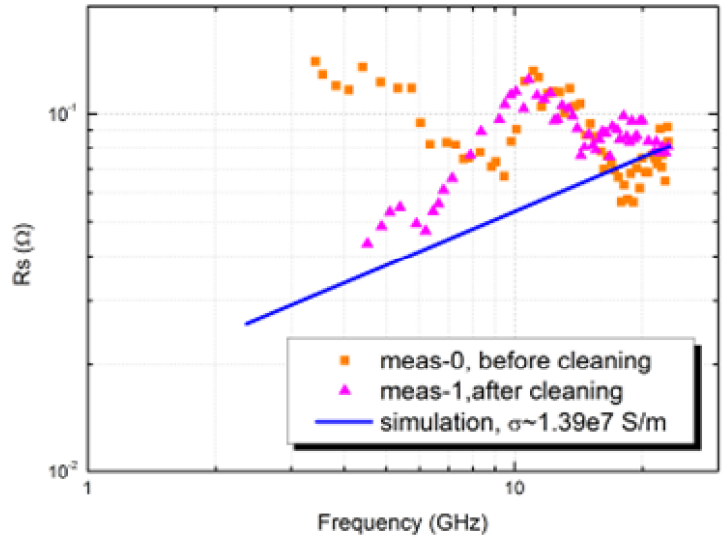

(b)

Figure 10: Report of (a) the Q-factor measurements and (b) the surface resistance compared with the simulation data.

From the Q-factor measurement of the empty cavity mode ($TM_{010}$) at 22GHz (where the central post is removed), the cavity loss is only due to the copper electric conductivity. Thus we can retrieve the electric metal conductivity value of about $1.39.10^7$ S/m for this cavity. We simulate the Q-factor and the surface resistance ($Rs$) versus frequency, corresponding to the blue curves in figure 10, which shows the usual square root dependence on frequency of losses. In figure 10, it is also evident for smaller gaps, i.e. frequencies below 8 GHz, an extra degradation of the Q-factor exists. This is mainly due to the lossy dielectric effect (copper oxidation), which is clearly shown to improve after cleaning. From 10 to

Corresponding author: Jean-Michel Le Floch; jeanmichel.lefloch@uwa.edu.au

15GHz, the drop in Q values is assumed to be mainly related to field leakage between the adjustable central post and the guiding hole from the cavity base (see in figure 7) as it has been already reported similar work, where a specially designed mechanism was implemented to maintain electrical connection with tuning [46]. To further improve these results over the tuning range this type of mechanism could be implemented.

**CONCLUSION**

In this work we have shown the validity of the LCR model and its limitations. Rigorous "in-house" finite element analysis has been implemented and allows accurate analysis at large gap sizes where the LCR model breaks down. Analysis was undertaken to understand the mode structure transformation from re-entrant mode to Transverse Magnetic mode during the removal process of the central post. The limitations on the LCR model was then determined and explained by the structure of the field changing. In practice we achieved gap sizes as low as 3μm and a solution for reducing machining roughness was discussed. For gap spacing of below 20μm equivalent to 1.4% of the total central post height, it was demonstrated that the LCR equivalent model could be used directly for both resonant frequency and Q-factor modeling, at larger gap spacing, for accuracy, a rigorous technique should be used. Based on the rigorous analysis a highly tunable capacity was made, which allowed tuning from 2 to 22 GHz. During the tuning process it was shown the re-entrant mode transformed to the standard $TM_{0,1,0}$ cavity mode, which was beyond the capabilities of the LCR model.

**ACKNOWLEDGMENT**

Part of this research results has been funded by the Australian Research Council, the French Research Agency (CNRS), with the partnership between France and Australia through the Australian Academy of Science ISL grant No. FR100027, and Le Conseil Régional du Limousin, cluster de calcul CALI and Labex Sigma-Lim (n° ANR-10-LABX-0074-01). This work was also supported by the Australian Research Council Grants FL0992016 and CE110001013. The authors would also like to thank Dr James Anstie for his Q-circle software used for extracting Q-factors from reflection measurements.

Corresponding author: Jean-Michel Le Floch; jeanmichel.lefloch@uwa.edu.au

Corresponding author: Jean-Michel Le Floch; jeanmichel.lefloch@uwa.edu.au

Corresponding author: Jean-Michel Le Floch; jeanmichel.lefloch@uwa.edu.au